\documentclass[jkps,preprint,fleqn,showpacs,showkeys]{revtex4}
\usepackage{graphicx}
\usepackage{amssymb}
\usepackage{amsmath}
\usepackage{bm}

\begin{document}
\title{ Is the Fano Antiresonance a Necessary Requirement for Circulating
        Currents in Mesoscopic Interferometers?}
\author{Yao Heng \surname{Su}}
\author{Sam Young \surname{Cho}}
\email{sycho@cqu.edu.cn}
\thanks{Fax: +86-23-65111531}
\author{Ai Min \surname{Chen}}
\affiliation{Centre for Modern Physics and Department of Physics,
Chongqing University, Chongqing 400030, China}

\author{Taeseung \surname{Choi}}
 \affiliation{Barom Liberal Arts College, Seoul Women's University, Seoul, 139-774, Korea}

\date{\today}

\begin{abstract}
 Coherent quantum tunneling effects on quantum interference are investigated
 in electron transport through a mesoscopic interferometer.
 An evanescent wave tunneling through a potential barrier in one arm can interfere
 with a propagating wave passing through the other arm of interferometer.
 It is shown that, even for the same arm lengths,
 such a quantum interference
 can induce a circulating current, where
 Fano antiresonances do not occur in electron transmission.
 It is found that there exists a critical value of asymmetric arm lengths that gives
 rise to a Fano antiresonance in electron transmission
 for the quantum interference between evanescent and propagating waves.
 We discuss the effects of Fano antiresonances originating from the
 asymmetric arm lengths on circulating currents.
\end{abstract}



\maketitle

\section{INTRODUCTION}

 Nonadditivity \cite{Webb,Washburn,Fuhrer} of parallel conductances is a prototype
 of quantum coherence in mesosocpic electronics \cite{mesoscopic}.
 It also implies that some classical circuit laws
 may not be valid due to nonlocality of electron waves in
 mesoscopic interferometers.
 As an another example,
 a recent theoretical study has shown that
 a circulating current \cite{Jayannavar} via a so-called current magnification
 and  negative current can flow along the loop path
 of interferometers owing to quantum interference even in the absence of
 magnetic fields, which cannot occur in classical parallel
 resistors. The quantum interference for the circulating current
 has been interpreted as
 a Fano-type interference \cite{Fano} with its characteristic transmission zeros \cite{Tekman},
  so called
 Fano antiresonance, in electron transport through interferometers only with
 an asymmetric arm lengths.

 The unique behaviors of circulating currents have been studied
 in various types of electronic interferometers such as
 single loop interferometers with a stub \cite{Cho1} or impurity
 potential \cite{Pareek,Vargiamidis} embedded in one of arms,
 an evanescent wave interferometer with a potential well \cite{Benjamin03},
 a multichannel interferometer with an impurity \cite{Bandopadhyay},
 double loop interferometers \cite{Yi,Park},
 double quantum dot interferometers \cite{Cho2,Cho3},
 multiple-arm interferometers \cite{Wu,Zhang},
 and spin-dependent interferometers \cite{Choi9802,Citro,Zhang}.
 In such interferometers,
 a circulating heat \cite{Cho3} and spin \cite{Choi9802,Citro,Zhang}
 currents have been reported as well as electric circulating currents.
 However, almost all such studies 
 have shown that
 the Fano antiresonances
 mediate such a circulating current in only geometrically asymmetric
 mesoscopic interferometers.
 Thus, it may be believed that the Fano antiresonances in electron transmission
 through mesoscopic interferometers are a
 necessary requirement for the existence of circulating currents.
 However,
 in an asymmetric multichannel interferometer
 with a potential impurity  \cite{Bandopadhyay},
 it was found that a circulating
 current can be induced without Fano antiresonances
 in a special energy range where there are two propagating
 modes with a bound state.

 It is not clear, to our knowledge, whether
 a Fano antiresonance
 is a necessary requirement for circulating currents
 in such two-terminal mesoscopic interferometers.
 To help our understanding on the question and to give
 a better understanding on fundamental quantum interference,
 in this study,
 we consider a realizable simplest model (see Fig. \ref{model}),
 that is a single loop interferometer with a potential barrier in
 each arms, which is enough to capture an essential physics.
 We focus on the quantum interference between propagating and
 evanescent waves respectively flowing through each arms of the interferometer.
 It is found that a circulating current can be induced by the
 quantum interference without transmission antiresonances
 even for the symmetric arm length of interferometer.
 For the circulating currents, the tunneling current through the potential barrier
 flows against the applied bias while a propagating current magnification
 occurs in the other arm.
 These behaviors of circulating currents persist up to a critical
 value of asymmetric arm lengths where electron transmission starts to have
 a Fano antiresonance.
 The effects of Fano antiresonances due to the asymmetric arm lengths
 are discussed on circulating currents.

\section{MODEL}

\begin{figure}
\includegraphics[width=9.0cm]{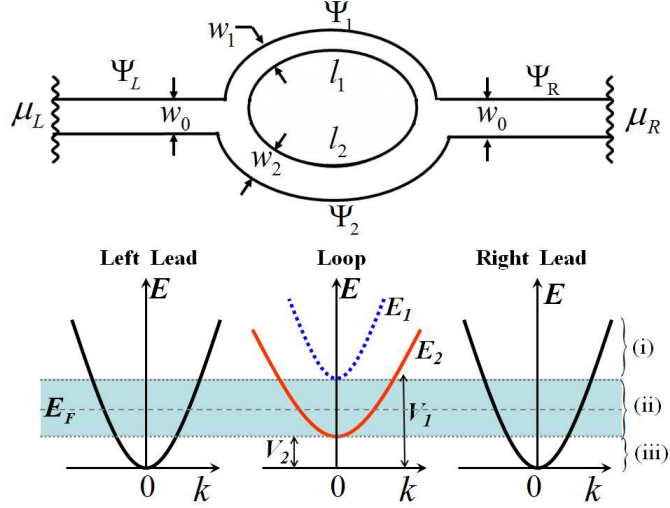}
 \caption{(color online)
 Top: Mesosocopic interferometers coupled to the left and right electron
 reservoirs that are respectively characterized
 by chemical potentials $\mu_L$ and $\mu_R$. $l_{1}$ and $l_{2}$
 denote the upper and lower arm length, respectively, such that the
 loop size of interferometer is $L=l_1+l_2$. $\Psi$'s are the
 electron wavefunctions in each regions.
 Bottom: Energy level diagram for each regions.
 If the loop has a narrower transverse width than the ideal leads,
 the energy bands $E_1$ and $E_2$ in the upper and lower arms  can be
 shifted up to have a higher energy than the energy bands in the leads,
 which can be modeled by introducing potential barriers $V_i$ ($i\in \{ 1, 2\}$)
 in each arms within one-dimensional waveguide theory.
 Here, the energies in the arms are depicted for $V_2 < V_1$.
 Quantum interferences in this interferometer can be characterized by
 three energy regimes. (i) For $0 < E_F < V_i$, electron can tunnel through
 each arms as an evanescent wave.
 (ii) For $V_i < E_F < V_j$ ($i\neq j$) in the shaded energy regime,
 electron propagates in one arm
 and tunnels through the other arm as an evanescent wave.
(iii) For $V_{j} <E_F$,
 electron propagates through the interferometer with different momentum $(V_1 \neq V_2$).
 This study is focused on the case (ii).}
 \label{model}
\end{figure}

 We start with an ideal loop interferometer with a single channel
 in the absence of magnetic fields
 in the top panel of Fig. \ref{model}. The interferometer is connected to two
 reservoirs where electron can be injected to the interferometer through
 the ideal leads $L$ and $R$
 when the chemical potentials are not equal each other, $\mu_R \neq \mu_L$.
 The geometrical symmetry of interferometer
 is determined by its upper and lower arm lengths $l_1$ and $l_2$.
 If the loop has a narrower transverse width than
 the ideal leads, an injected electron momenta cannot be matched
 with the electron energy in the loop arms due to their narrower transverse confinement.
 Thus, in the arm, electron undergoes
 a decaying of wavefunction, {\it i.e.}, an evanescent wave.
 This phenomena can be well described by introducing arm potentials
  $V_1$ and $V_2$ respectively for the upper and lower arms
 within one-dimensional waveguide theory \cite{Xia}.
 These potentials can make the energy levels of the upper and
 lower arms higher than those of the ideal leads.
 For $V_2 < V_1$, the energy level diagrams are depicted in the bottom panel of Fig.
 \ref{model}.

 We concentrate on the energy region (ii) $V_2 <
 E_F < V_1$ in the bottom panel of Fig. \ref{model},
 where the quantum interference between propagating and
 evanescent waves occurs.
 We employ one-dimensional waveguide theory \cite{Xia} with local coordinates.
 The Hamiltonians are given as
 $H_{L,R}=p^2/2m^*$ for the left and right leads and
 $H_{1,2}=p^2/2m^*+V_{1,2}$ for the upper and lower arms,
 where $p$ is the electron momentum, $m^*$ is
 the electron effective mass, and $V_{1,2}$ are the electric potentials
 in the upper and lower arms, respectively.
 Suppose that an electron is injected from the left lead with its energy
 $E_F$.
 The wavefunctions for each regions
 are respectively written as
 $\Psi_L(x) = e^{ik_Fx} + r e^{-ik_Fx}$ for the left lead,
 $\Psi_1(x) = a e^{-\kappa_F x}+b e^{\kappa_F x}$ for the upper arm,
 $\Psi_2(x) = c e^{ik'_F x}+d e^{-ik'_F x}$ for the lower arm, and
 $\Psi_R(x) = t e^{ik_Fx}$ for the right lead,
 where $k_F = \sqrt{2m^* E_F}/\hbar$,
 $\kappa_F =\sqrt{K_1^2-k^2_F}$, and
 $k'_F=\sqrt{k^2_F-K_2^2}$ with $K_i=\sqrt{2m^*V_i}/\hbar$ ($i \in \{1,2\}$).
 By using the Griffith boundary conditions \cite{Griffith}
 (current conservation)
 and the continuities of wavefunctions
 at the left and right junctions, one can
 obtain the coefficients of wavefunctions $r$, $a$, $b$, $c$, $d$, and $t$
 in terms of $k_F$, $\kappa_F$, $k'_F$, $l_1$, and $l_2$.

 The probability current densities $J_j(x) =
 \frac{e\hbar}{2mi}\left(\Psi_j^*(x) \partial_x \Psi_j(x)-\Psi_j(x)
 \partial_x \Psi_j^*(x)\right)$ ($j \in \{L, 1, 2, R\}$)
 through the whole interferometer
 and the upper and lower arms can be respectively obtained as
\begin{subequations}
 \begin{eqnarray}
 J &=&J_0 \, {\cal F}_0
      \left({\cal F}_1-{\cal F}_2\right)^2,
 \label{J} \\
 J_{1}&=&J_0 \, {\cal F}_0
      \left({\cal F}_1-{\cal F}_2\right){\cal F}_1,
 \label{J1}\\
 J_{2}&=&J_0 \, {\cal F}_0
      \left({\cal F}_2-{\cal F}_1\right){\cal F}_2,
  \label{J2}
 \end{eqnarray}
\end{subequations}
where $J_0 = e\hbar k_F/m^*$ is the unit probability current
density. Here,
 ${\cal F}_0 = 4 k^{2}_F\kappa^2_F k'^2_F /[4k^{2}_F\big(
  \kappa_F\cosh\kappa_F l_1 \sin k'_Fl_{2}+k'_F\sinh \kappa_F l_1 \cos k'_Fl_{2} \big)^2
 \!\! +\! \big(\! ( k_F^2\!-\kappa^2_F + k'^2_F ) \sinh \kappa_F l_1 \sin k'_Fl_{2}
  + 2\kappa_F k'_F (1\!-\!\cosh \kappa_F l_1 \cos k'_Fl_2 )\big)^2]$,
   $k'_F {\cal F}_1=\sin k'_Fl_{2}$, and $\kappa_F {\cal F}_2=-\sinh\kappa_F l_{1} $.
 The transmission amplitude is also written as
 $T = J/J_0 = {\cal F}_0 ({\cal F}_1 - {\cal F}_2)^2$.

\section{CIRCULATING CURRENT AND ITS GENERATING CONDITION}

 A transport current $J$ flows from the left to the right,
 as assumed $\mu_R < \mu_L$,
 and is split into two local currents $J_1$ and $J_2$
 respectively
 flowing in the upper and lower arms under the current conservation.
 If $J_1$ ($J_2$) is bigger than
 $J$, $J_2$ ($J_1$) should flow against applied bias due to the
 current conservation \cite{Jayannavar}. These phenomenon can be
 understood by introducing a circulating current $J_c$ flowing along
 the loop path of interferometer. Consequently, the arguments of
 circulating current can be summarized by the expression \cite{Yi}
 of circulating current
 \begin{equation}
 J_c =\frac{\textrm{sign}[J_1]}{2}\Big(|J_1|+|J_2|-J\Big).
 \label{definition}
 \end{equation}
 Equation (\ref{definition}) shows that
 any classical parallel resistor cannot have a circulating current.
 The definition then allows to capture a pure quantum mechanical effect
 for electron transport through two-terminal interferometer.

 According to the arguments of circulating current
 with the current expressions in Eqs. (\ref{J})-(\ref{J2}),
 one can obtain a generating condition of circulating current
 as a function of parameters $\{K_1L,K_2L,\beta,k_FL\}$,
 where $\beta=l_1/l_2$ characterizes the geometrical asymmetry
 of interferometer.
 $\beta=1$ denotes a geometrically symmetric interferometer.
 If $J < J_1$ and $J_2 < 0$, one can assign
 a circulating current by the generating condition written
 as
\begin{subequations}
 \begin{equation}
  {\cal F}_1(K_2L,\beta,k_FL) <  {\cal F}_2 (K_1L,\beta,k_FL) < 0.
  \label{condition1}
 \end{equation}
 From  Eq. (\ref{definition}),
 circulating currents satisfying Eq. (\ref{condition1})
 flow in \textit{clockwise} direction.
 If $J < J_2$ and $J_1 < 0$,
 the generating condition for \textit{counterclockwise} circulating current
 is given as
 \begin{equation}
  {\cal F}_2 (K_1L,\beta,k_FL) <  {\cal F}_1(K_2L,\beta,k_FL)  < 0.
  \label{condition2}
 \end{equation}
\end{subequations}

\section{CIRCULATING CURRENT WITHOUT FANO ANTIRESONANCES FOR SYMMETRIC ARM LENGTH}

\begin{figure}
\includegraphics[width=6.0cm]{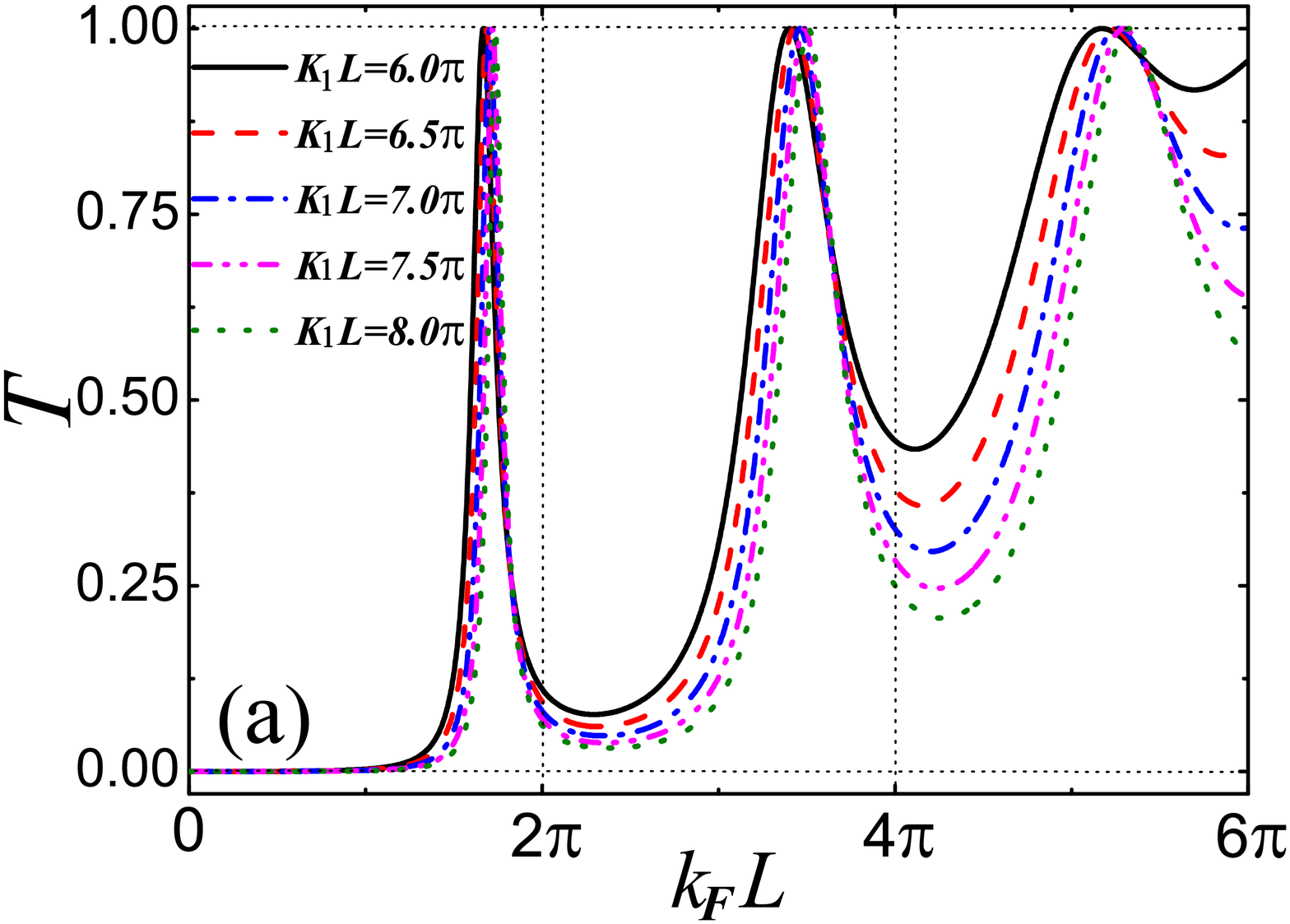}
\includegraphics[width=6.0cm]{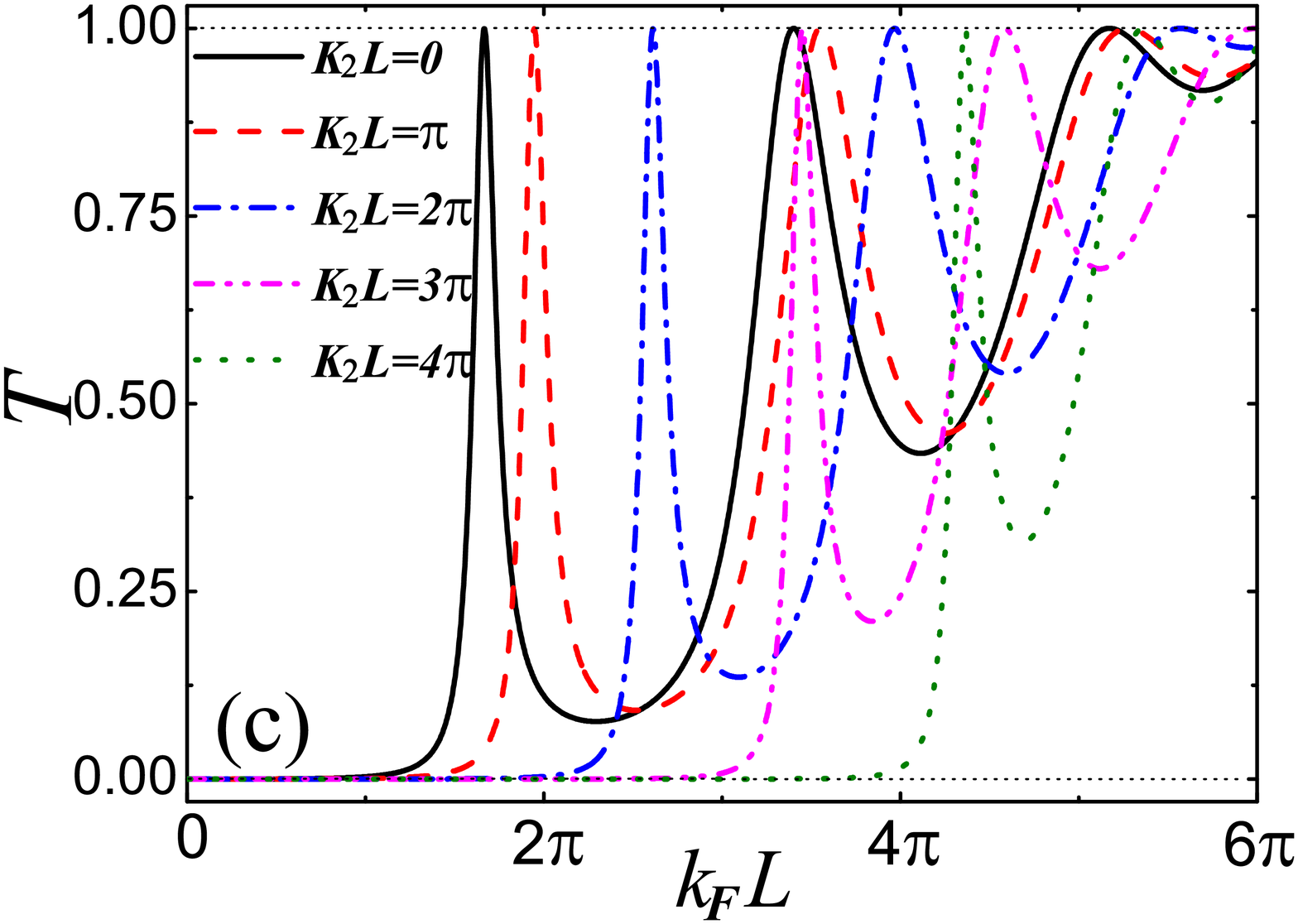}
\includegraphics[width=6.0cm]{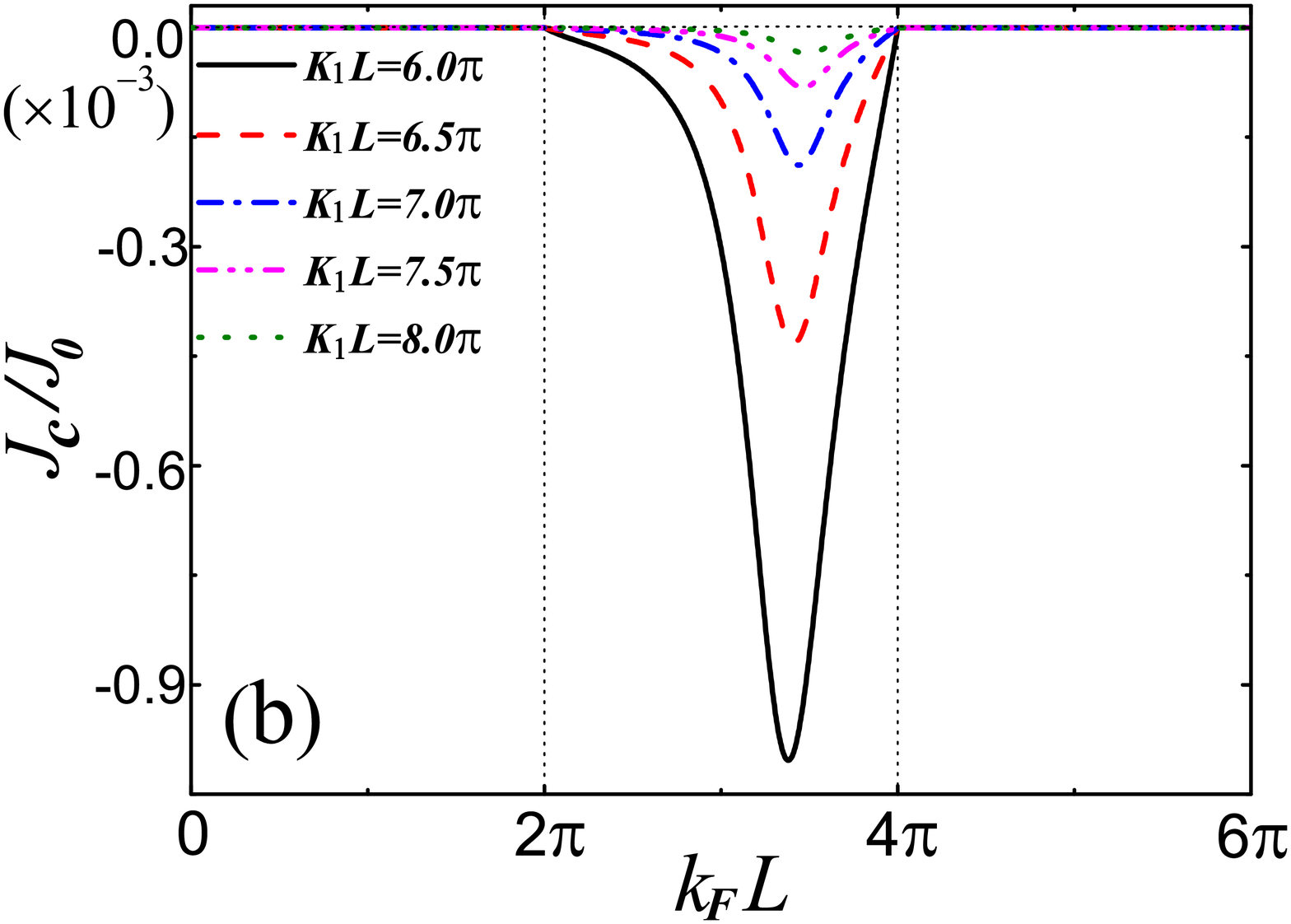}
\includegraphics[width=6.0cm]{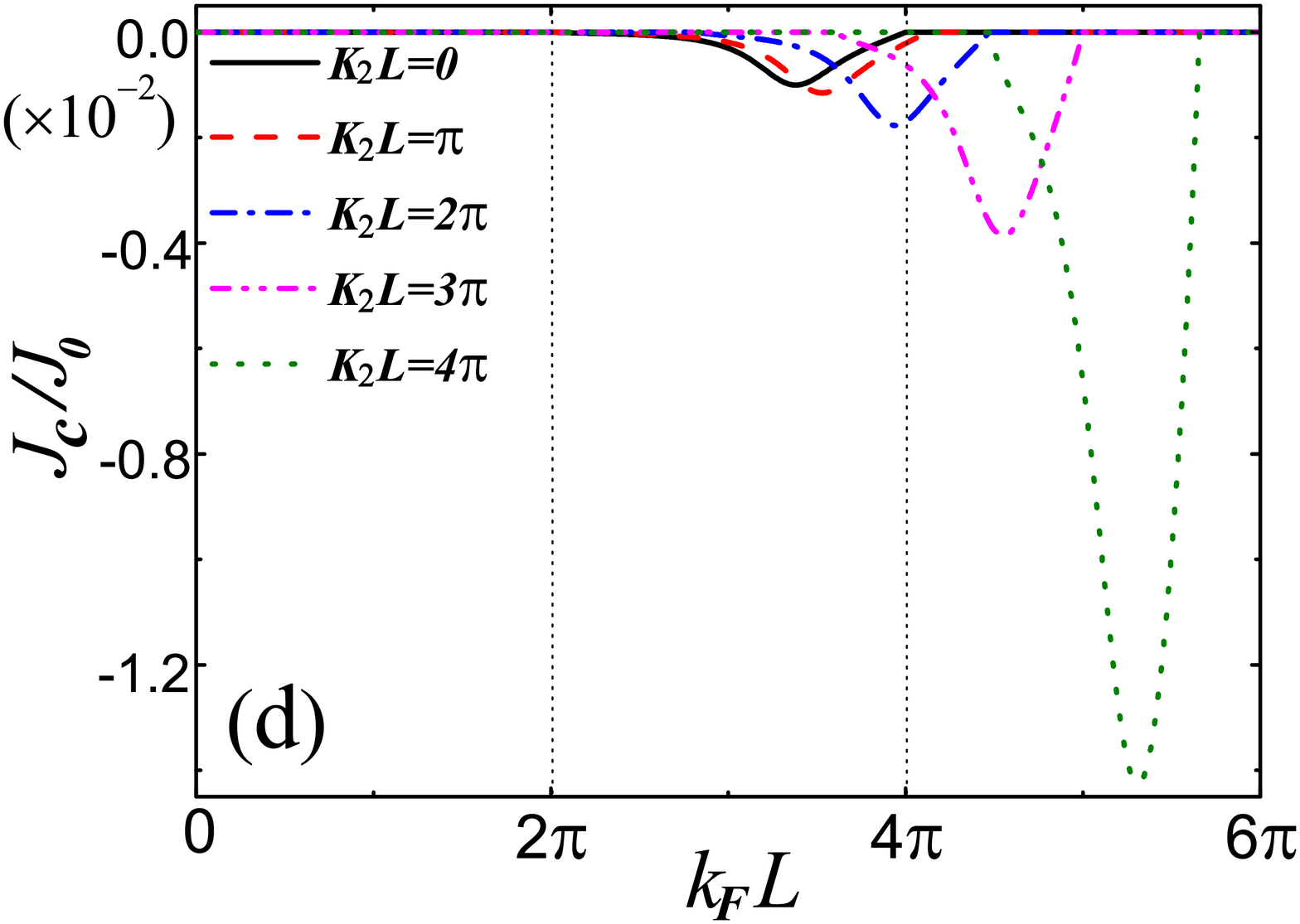}
 \caption{(color online)
   Transmission amplitudes $T$ and circulating currents $J_c/J_0$
   as a function of  $k_FL$ for the
   symmetric arm length $l_1=l_2$.
   Here, $J_0 = e\hbar k_F/m^*$ and $K_i=\sqrt{2m^*V_i}/\hbar$ ($i\in \{1,2\}$).
   Note that no transmission zeros are seen in (a) and (c).
   Left:
   (a) $T=J/J_0$ and (b) $J_c/J_0$
   for various values of $K_1L$ and $K_2L=0$.
   As $K_1L$ increases from $K_1L=6 \pi$, a smaller circulating current flows
   in the same range $(2\pi ,4\pi)$ of $k_FL$.
   Right: (c) $T=J/J_0$ and (d) $J_c/J_0$
   for $K_1L=6 \pi$ and various values of $K_2L$.
   As $K_2L$ increases with $K_1L=6\pi$,
   a larger circulating current is seen in the shifted ranges
   $(\sqrt{(2\pi)^2+(K_2L)^2} ,\sqrt{ (4\pi)^2+(K_2L)^2})$ of $kL$ for the chosen parameters.
   If one choose a value of $K_2L$ in the range $\sqrt{32}\, \pi < K_2 L <
   K_1L=6\pi$, no circulating current can exist. }
 \label{fig2}
\end{figure}

 Let us consider first the
 symmetric arm length $l_1=l_2$ for $K_1L \neq 0 $ and $K_2L=0$.
 We plot the transmission amplitudes $T=J/J_0$ and
 circulating currents $J_c/J_0$ as a function of $k_FL$ in Figs. \ref{fig2} (a) and (b),
 respectively. Note that no transmission zeros are seen in Fig. \ref{fig2} (a).
 However, Fig. \ref{fig2} (b) shows that there exist circulating
 currents even in the geometrically symmetric interferometers.
 Also, our circulating current shows its very different behaviors from the
 ones of geometrically asymmetric interferometers, showing that
 Fano antiresonances in electron transmission
 accompany a circulating current,
 in Refs. \cite{Jayannavar,Pareek,Cho1,Vargiamidis,Benjamin03,Wu,Zhang,Choi9802,Citro}.

 The characteristic properties of
 our circulating current in geometrically symmetric interferometer
 for the quantum interference between propagating and evanescent waves
 are summarized as follows:
 (i) From
   ${\cal F}_2 \lneqq {\cal F}_1$ for $\beta=1$ and
   ${\cal F}_2 < {\cal F}_1 < 0$ in Eq. (\ref{condition2}),
   the generating condition of circulating current becomes ${\cal
   F}_1 < 0$.
  Thus, a circulating current exists in the regions $(2(2 m-1)\pi, 4m\pi)$
 of $k_FL$ for $4m\pi \leq K_1L \leq 2(2m+1)\pi$
 or in the regions $(2(2 m-1)\pi, K_1L)$ of $k_FL$ for $2(2 m-1)\pi \leq K_1L \leq
 4m\pi$. Here,
 $m$, a positive integer, corresponds to
 the $m$-th transmission resonance for the range $k_FL((2 m-1)\pi, 2m\pi)$.
 Hence, the $2m$-th transmission resonance accompany a circulating
 current.
 This interesting behavior can be understood by a following way.
 Suppose that there is an isolated loop with a high potential which forms
 a quantum well in the loop. For our interferometer,
 the eigen energies of the quantum well in the isolated loop may
 roughly
 one-to-one correspond to the resonant transmission energies
  $E_m \sim \alpha \frac{\hbar^2}{2m^*L^2} (2m\pi)^2$,
 where a parameter $\alpha$ is responsible for the effect of coupling to the leads
 and must be smaller than $1$ $(0 < \alpha < 1)$.
 If the higher potential $V_1$ (here, $K_1L$)  becomes bigger, the resonant energies
 become closer to the eigen energies of the quantum well, which is
 shown in Fig. \ref{fig2} (a).
 Also, in the coordinate of isolated loop,
 the $2m$-th eigen state of the quantum well has a odd parity
 wavefucntion while the $(2m-1)$-th eigen state has a even parity
 wavefunction. Then, our circulating currents may be mediated by
 the odd parity (anti-symmetric) wavefunctions of the quantum well.
 (ii) A tunneling current $J_1$ flows only against the applied bias in the regions.
 Compared to the circulating current in
 the previous studies of geometrically asymmetric interferometers,
 our circulating current can flow only in one direction, {\it i.e.}, electron circulates
 in counterclockwise (clockwise) direction for $V_1 > V_2$ $(V_1 < V_2)$
 in the geometrically symmetric interferometer $(l_1=l_2)$.
 In other words, the asymmetric potential barriers in the interferometer determines
 the direction of circulating current.
 (iii) A bigger propagating current $J_2$ than transport current $J$ flows
       in the other arm.
  (iv) The amplitude of circulating currents decreases as $K_1L$ increases
  in a given range of $k_FL$ where a circulating current can exist.

 For $K_2 L\neq 0$ and $K_1L=6\pi$,
 we display the transmission amplitudes $T=J/J_0$ and
 circulating currents $J_c/J_0$ as a function of $kL$ in Fig. \ref{fig2} (c) and (d),
 respectively.
 Figure \ref{fig2} (d) shows that
 no circulating current is induced when two evanescent waves
 interfere for $0 < E_F < V_2 < V_1$.
 When propagating and evanescent waves interfere for $V_2 < E_F <
 V_1$,
 as $V_2$ increases up to $K_2 L = \sqrt{32}\, \pi$ from zero,
 the region of circulating current changes to a shifted and narrower region
 $(\sqrt{(2\pi)^2+(K_2L)^2} ,\sqrt{ (4\pi)^2+(K_2L)^2})$ from
 $(2\pi, 4\pi)$ in $k_FL$ and a larger circulating current can flow.
 However, it should be noted that as $V_2$ increases further to the value of
 $\sqrt{32}\, \pi < K_2 L < K_1L=6\pi$, circulating current cannot be
 induced.

\section{A CRITICAL VALUE OF ASYMMETRIC ARM LENGTHS AND FANO ANTIRESONANCES}

 Now, let us discuss about Fano antiresonances and their
 effects on circulating currents in the geometrically asymmetric interferometer $(l_1\neq l_2)$.
 From Eq. (\ref{J}), one finds the
 antiresonance condition given as
\begin{equation}
   {\cal F}_1(K_2L,\beta,k_FL)= {\cal F}_2 (K_1L,\beta,k_FL).
 \label{anti}
\end{equation}
 At the antiresonances, from Eqs. (\ref{J})-(\ref{J2}), and (\ref{anti}),
 the circulating currents also vanish as well as the transport currents.
 For a given $K_1L$ and $ K_2L$,
 there can be
 $s$ solution sets $\{k^{(s)}_FL\}$
 satisfying Eq. (\ref{anti}) at a certain value $\beta^{(s)}$,
 where $s$ is a positive integer,
 which implies that $s$ Fano antiresonances occur in electron transmission.

\begin{figure}
\includegraphics[width=6.0cm]{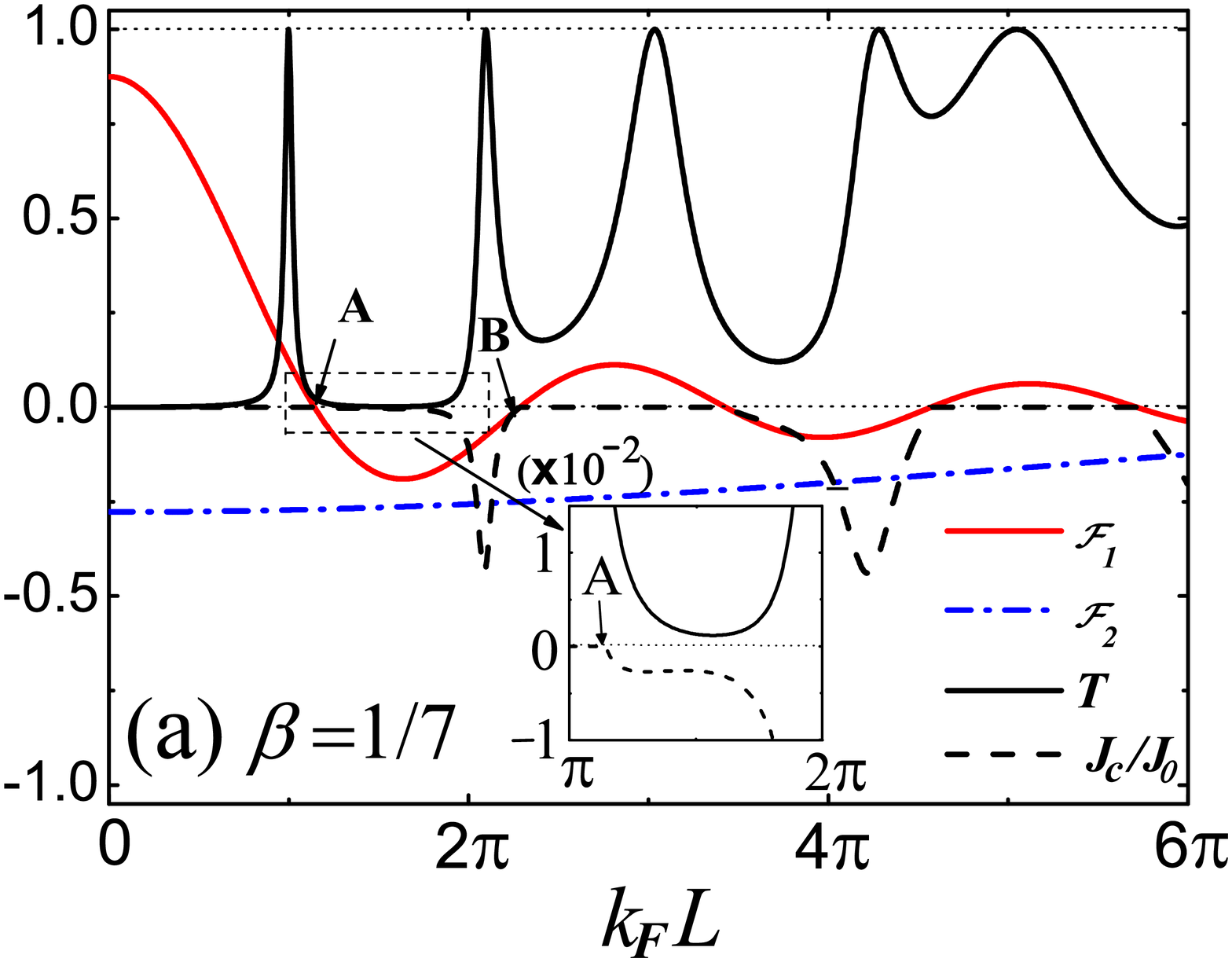}
\includegraphics[width=6.0cm]{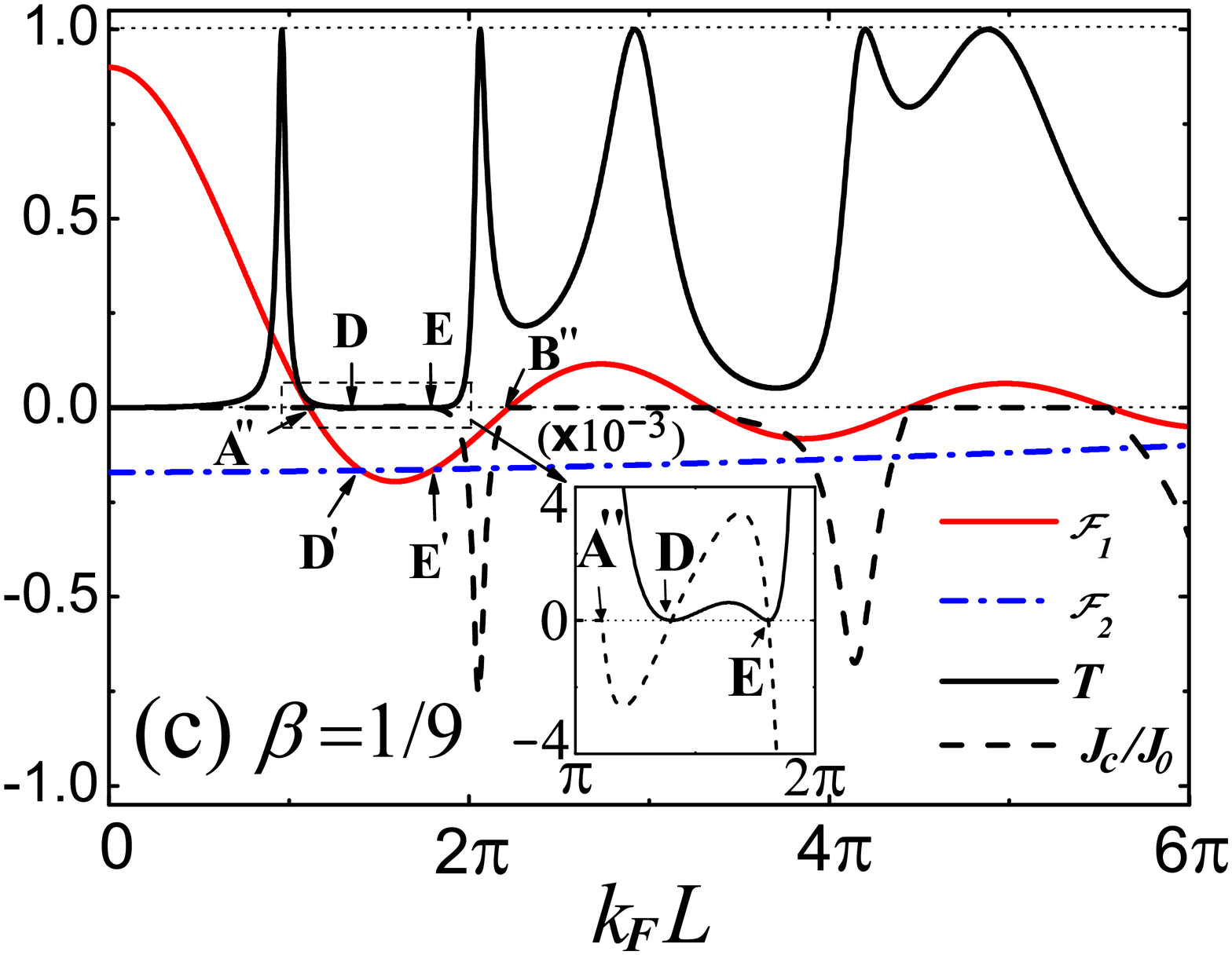}
\includegraphics[width=6.0cm]{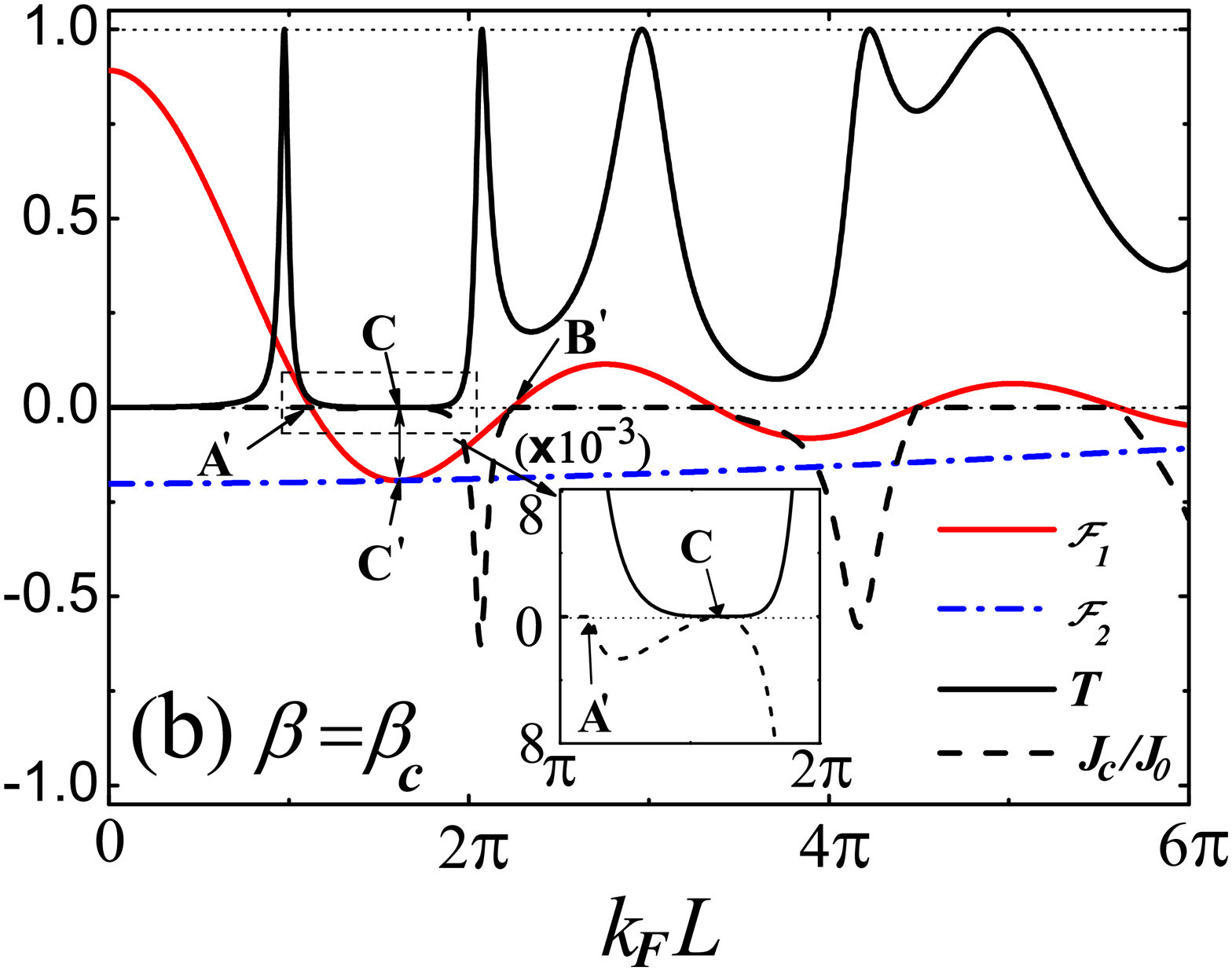}
\includegraphics[width=6.0cm]{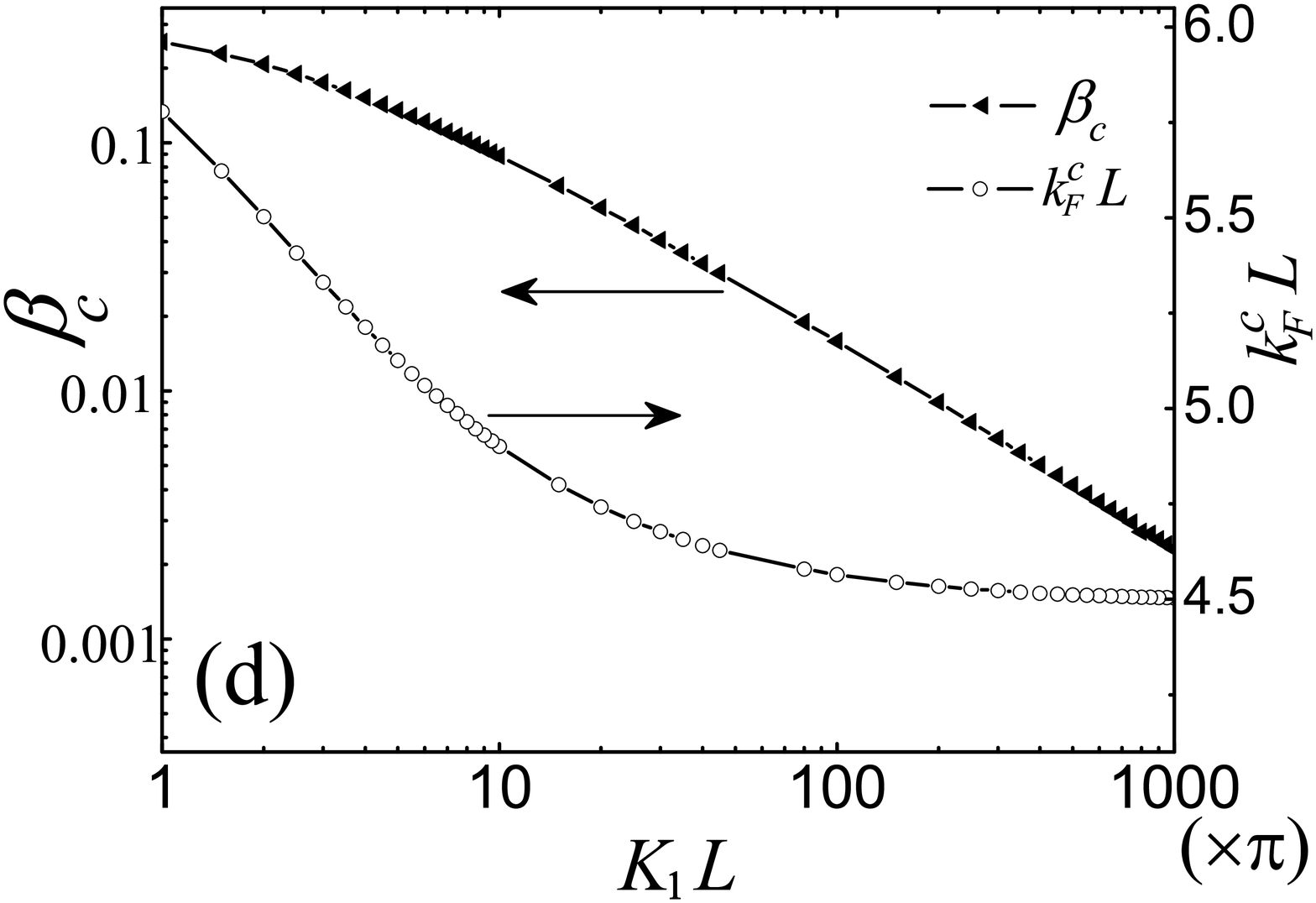}
\caption{ (color online)
    Transmission amplitudes $T$ (black solid lines),
   circulating currents $J_c/J_0$ (black dashed lines), and
   the characteristic functions ${\cal F}_1$ (red solid lines)
   and ${\cal F}_2$ (blue dash-dotted lines)
   as a function of  $k_FL$
   for (a) $\beta (=l_1/l_2)=1/7$ (no Fano antiresonance),
   (b) $\beta = \beta_c \simeq 0.121704$ (one Fano antiresonance),
   and (c) $\beta=1/9$ (two Fano antiresonances)
   in the interferometer with an asymmetric arm length $l_1 \neq l_2$.
   Here, $K_1L=6 \pi$ and $K_2L = 0$.
   For ${\cal F}_1=0$, $J_c=0$, e.g., here at
   (a) $A$ and $B$ of $k_F L$, (b) $A'$ and $B'$ of $k_FL$, and  (c) $A''$ and $B''$ of $k_FL$.
   At these $k_FL$, $T \neq 0$.
   For ${\cal F}_2 < {\cal F}_1 < 0$, $J_c < 0$ and,
    for ${\cal F}_1 < {\cal F}_2 < 0$, $J_c > 0$.
   For ${\cal F}_1={\cal F}_2$, the Fano antiresonances ($T=0$) occur
   and $J_c=0$ at the points $C$ in (b) and
    $D$ and $E$ in (c),
   which are shown from the pictorial representations of solutions.
   (d) Critical values $\beta=\beta_c$ and $k_FL=k^c_FL$ as a function of  $K_1L$ with $K_2L=0$,
   where only one Fano antiresonance occurs in electron transmission.
     }
  \label{fig3}
\end{figure}

 For instance, in Fig. \ref{fig3},
 we consider the case of $K_1L=6\pi$ and $K_2L=0$
 and plot $T$, $J_c/J_0$, ${\cal F}_1$, and ${\cal F}_2$ as a function of $k_FL$
 for various asymmetric arm lengths (a) $\beta=1/7$,
 (b) $\beta=\beta_c\simeq 0.121704$,
 and (c) $\beta=1/9$.
 In $k_FL$,
 there are the periodic existing ranges of circulating currents
 from the generating condition Eqs. (\ref{condition1}) and (\ref{condition2})
 since
 ${\cal F}_1$ is a sinusoidal function of $k_FL$ while ${\cal F}_2$
 is a negative monotonic function of $k_FL$.
 (i) For $\beta > \beta^{(1)}=\beta_c$, no Fano antiresonance occurs because
 ${\cal F}_2 < {\cal F}_1$ in $k_FL$ (Fig. \ref{fig3} (a)).
 The existing ranges of circulating currents are determined by ${\cal F}_1(\beta) < 0$.
 The circulating currents flows in the counterclockwise direction.
 (ii)
 For $\beta=\beta_c$, eventually, a Fano antiresonance appears at $k^c_FL$.
 Note that $k^c_FL$ ($C$ in Fig. \ref{fig3} (b)) locates inside between the first existing range
 $k_FL(A',B')$
 determined by ${\cal F}_1(\beta_c) < 0$. Thus, when $k_FL$  is
 swept across the antiresonance,
 the direction of circulating current does not changed and
 the circulating current just disappear at $k^c_FL$.
 (iii) As $\beta$ decreases further for $ \beta^{(3)} <\beta < \beta^{(1)}$,
 the electron transmission has two Fano antiresonances
 inside between the first existing range
 determined by ${\cal F}_1(\beta^{(3)} <\beta < \beta^{(1)}) < 0$.
 Figure \ref{fig3} (c) shows that the circulating current direction
 is changed to the opposite (clockwise) direction in $k_FL(D,E)$ between the pair of
 antiresonances.
 As the ordered pairs of antiresonances is split to
 approach closer to the transmission resonances by decreasing further smaller $\beta$,
 a larger circulating current flows.

 For $\beta^{(2s+1)} < \beta < \beta^{(2s-1)}$, $2s$ Fano
 antiresonances occur and the circulating current direction
 in $k_FL$ between the ordered pairs of
 antiresonances
 is changed to the opposite direction.
 Note that the pairs of antiresonances exist inside between the
 existing ranges of circulating currents determined by ${\cal F}_1(\beta) < 0$.
 Therefore,
 the $2m$-th transmission resonances play a significant role for a circulating
 current even though Fano antiresonances occur in between the ordered pairs of
 antiresonances of
 electron transmission where the circulating current direction
 is changed to the clockwise direction from the counterclockwise direction.

 Figure  \ref{fig3} (d) shows that Fano antiresonances can occur in electron
 transmission through the interferometer when
 the arm length of propagating wave is longer than that of evanescent
 wave, {\it i.e.}, $ l_1 < l_2$,
 As $K_1L$ increases, $\beta_c$ decreases exponentially and $k^c_FL$ approaches to
 about $4.5$.

\section{CONCLUSIONS}

 Coherent quantum tunneling effects on quantum
 interference in a mesoscopic interferometer have been investigated to
 address whether a circulating current does require a Fano antiresonance in
 electron transmission.
 It is found that
 the quantum interference between the propagating and evanescent
 waves can induce a circulating current without transmission zeros
 even for the symmetric arm length of interferometer.
 A Fano antiresonance
 of electron transmission was shown to start appearing at a critical
value of asymmetric arm lengths. As a result,
 it was shown that Fano antiresonance is not
 a necessary requirement for circulating currents in two-terminal
 mesoscopic interferometers.

\begin{acknowledgments}
 We thank Huan-Qiang Zhou for helpful discussions.
 This work was supported by the NSFC under Grant No.10874252
 and Natural Science Foundation  Project of CQ CSTC.
 T. Choi was supported by a research grant from Seoul Women's
 University(2009).
\end{acknowledgments}


\end{document}